\documentclass[aps,prl,twocolumn,superscriptaddress,notitlepage,nofootinbib]{revtex4-1}
\usepackage{amsmath}
\usepackage{bm}
\usepackage{graphicx}
\usepackage{color,soul}
\usepackage{float}
\usepackage{multirow}
\usepackage[inline]{enumitem}

\def\bea#1\eea{\begin{align}#1\end{align}}
\newcommand{\nn}{\nonumber\\}
\newcommand{\bef}{\begin{figure}[h!tb]\centering}
\newcommand{\eef}{\end{figure}}

\allowdisplaybreaks

\begin{document}
\title{Threshold Resummation for Hadron Production in the Small-$x$ Region}

\author{Hao-Yu Liu }
\affiliation{Center of Advanced Quantum Studies, Department of Physics, Beijing Normal University, Beijing 100875, China}

\author{Zhong-Bo Kang }
\email{zkang@physics.ucla.edu}
\affiliation{Department of Physics and Astronomy, University of California, Los Angeles, California 90095, USA}
\affiliation{Mani L. Bhaumik Institute for Theoretical Physics, University of California, Los Angeles, California 90095, USA}

\author{Xiaohui Liu }
\email{xhliu@bnu.edu.cn}
\affiliation{Center of Advanced Quantum Studies, Department of Physics, Beijing Normal University, Beijing 100875, China}

\date{\today}

\begin{abstract}
We study the single hadron inclusive production in the forward rapidity region in proton-nucleus collisions. We find the long-standing negative cross section at next-to-leading-order (NLO) is driven by the large negative threshold logarithmic contributions. We established a factorization theorem for resumming these logarithms with systematically improvable accuracy within the color glass condensate formalism. We demonstrate how the threshold leading logarithmic accuracy can be realized by a suitable scale choice in the NLO results. The NLO spectrums with the threshold logarithms resummed remain positive and impressive agreements with experimental data are observed. 

\end{abstract}

\date{\today}

\maketitle

%%%%%%%%%%
{\it Introduction}. 
Gluon saturation has attracted a lot of attention in recent years in nuclear physics community. This is in particularly true during the rapid development towards the realization of the Electron Ion Collider (EIC), where one of the scientific goals is to search for gluon saturation and to explore the properties of such a regime~\cite{Accardi:2012qut,NAS-EIC}. Gluon saturation plays the key role in understanding proton and heavy nuclei collisions in the high energy limit, where the gluon momentum fraction $x$ is very small. In such a  small-$x$ region, the gluon density grows dramatically and enters the nonlinear regime where the gluon recombination becomes equally important to the splitting, and the Color Glass Condensate (CGC) effective theory~\cite{Gelis:2010nm,Mueller:1989st,Mueller:1993rr} is the proper framework to describe such a regime. The nonlinear B-JIMWLK equation~\cite{Balitsky:1995ub, Kovchegov:1999yj, JalilianMarian:1997jx, JalilianMarian:1997gr, Iancu:2000hn, Ferreiro:2001qy} replaces the position of the linear BFKL equation~\cite{Balitsky:1978ic}, which inevitably leads to the gluon saturation~\cite{Gribov:1984tu,Mueller:1985wy} with a characteristic scale $Q_s$. The saturation scale $Q_s$ features the typical transverse momentum of the gluons inside the proton or the nucleus and grows as $x$ decreases. 

Experimental efforts have been made to identify the saturation phenomenon. Earlier experimental hints on gluon saturation include extensive measurements on structure function in deep inelastic scattering at HERA~\cite{GolecBiernat:1998js}, and the strong suppression of both single hadron~\cite{Arsene:2004ux,Adler:2004eh,Adams:2006uz} and dihadron production~\cite{Adler:2005ad,Adler:2005ad,Braidot:2010ig} cross sections at forward rapidity in d+Au collisions at the Relativistic Heavy Ion Collider (RHIC). More recently the measurements at the Large Hadron Collider~\cite{Acharya:2018hzf,Aad:2016zif,Dusling:2013oia} are also compatible with the saturation-model predictions. In the future, the dedicated measurements at the future EIC will provide further information on gluon saturation. 

In order to faithfully and unambiguously establish gluon saturation and its onset, reliable theoretical predictions for the small-$x$ phenomena at colliders are crucial. When $Q_s \gg \Lambda_{\rm QCD}$ and thus the coupling constant $\alpha_s(Q_s)\ll 1$, the theoretical predictions can be built upon perturbative QCD with a suitable factorization framework. However, for the semi-hard saturation scale of a few GeVs, $\alpha_s(Q_s)$ is typically not small enough. As a consequence, calculations beyond the leading order (LO) are generally required to ensure the convergence of the perturbative results. Recently, tremendous progress have been made in realizing the next-to-leading order (NLO) calculations for the small-$x$ physics~\cite{Dumitru:2005gt,Altinoluk:2011qy,Chirilli:2011km,Beuf:2011xd,Beuf:2016wdz,Beuf:2017bpd,Boussarie:2016ogo,Boussarie:2016bkq,Balitsky:2012bs,Roy:2018jxq,Liu:2019iml,Roy:2019cux,Roy:2019hwr}. 

In the physical processes investigated so far, single inclusive hadron production in proton-nucleus collisions, $pA\to hX$, is among the most studied ones. This will be the main focus of our current paper. The seminal work~\cite{Chirilli:2011km} confirms the CGC factorization for this observable at the NLO. However, the exhibited negative cross section when the hadron transverse momentum $p_{h,\perp}$ becomes a bit larger was quite a puzzle in the community~\cite{Stasto:2013cha}. Significant efforts have been devoted to resolve this issue, see e.g.~\cite{Kang:2014lha,Altinoluk:2014eka,Watanabe:2015tja,Ducloue:2016shw,Iancu:2016vyg,Xiao:2018zxf,Liu:2019iml} and references therein. In one of the most recent works~\cite{Liu:2019iml}, the approach introduced can maintain the positivity of the cross section to medium $p_{h,\perp}$ region. However, the cross section eventually becomes negative for even larger $p_{h,\perp}$, although such a transverse momentum is perfectly allowed with $p_{h,\perp} \ll \sqrt{s}$. It is thus widely accepted that the practical phenomenological applications of the NLO calculations for this process are by far problematic~\cite{Mantysaari:2019nnt,Mantysaari:2020axf}.

In this work, we present solid evidence that the threshold logarithm in the QCD perturbation series is the source to the negative cross section. We are able to resum these logarithms to all orders at the leading logarithmic accuracy (${\rm LL}_{\rm thr.}$). We find that 
after resummation, %the probability of the real soft emission is suppressed and 
the NLO predictions with the threshold logarithms resummed (${\rm NLO} + {\rm LL}_{\rm thr.}$) stay positive and agree well with the experimental data. %meanwhile the theoretical uncertainties get dramatically reduced. 
Early suggestion of such logarithms as solutions to the negative spectrum problem can be found in~\cite{Xiao:2018zxf,Kang:2019ysm}. In the same spirit, it might be interesting and instructive to notice that collinear logarithms in the NLO BK equation is the main source responsible for the unstable or even negative solutions and an improved equation with these collinear logarithms resummed solves this instability~\cite{Beuf:2014uia,Iancu:2015vea,Iancu:2015joa,Lappi:2016fmu,Hatta:2016ujq,Ducloue:2019ezk}.

{\it Threshold logarithms.} 
Threshold logarithms are common features of the partonic cross sections for hadronic processes~\cite{Sterman:1986aj,Becher:2007ty,Aicher:2010cb}. They are expected to be large and therefore invalidate the truncations in the perturbative expansion in $\alpha_s$, when a massive final state is produced or kinematic constrains are implemented to force the system reaching its maximally allowed energy. Even in cases where all the kinematics are away from the machine threshold, such as the $125\>{\rm GeV}$ Higgs production at the $13\>{\rm TeV}$ LHC, the threshold logarithms are still found to be sizable~\cite{deFlorian:2016spz}, due to the steep falling shape of the parton distribution functions (PDFs)~\cite{Becher:2007ty}, which effectively restricts the maximally allowed energy and enhances the effects. 
Conventional wisdom to rescue the perturbative predictive power is to resum the threshold logarithms $L$~\cite{Sterman:1986aj,Becher:2007ty,Aicher:2010cb}, which formatively turns the fixed order (FO) series 
$\sum_n^{\rm FO} \alpha_s^n (\sum_k L^k + c_n)  \to 
e^{g(L)}
\sum_n^{\rm FO } \alpha_s^n c_n
$, where $\sum_n^{\rm FO } \alpha_s^n c_n$ is free of large corrections and a fixed order truncation is therefore justified. 

\begin{figure}[H]
  \centering
\includegraphics[width=2.5in]{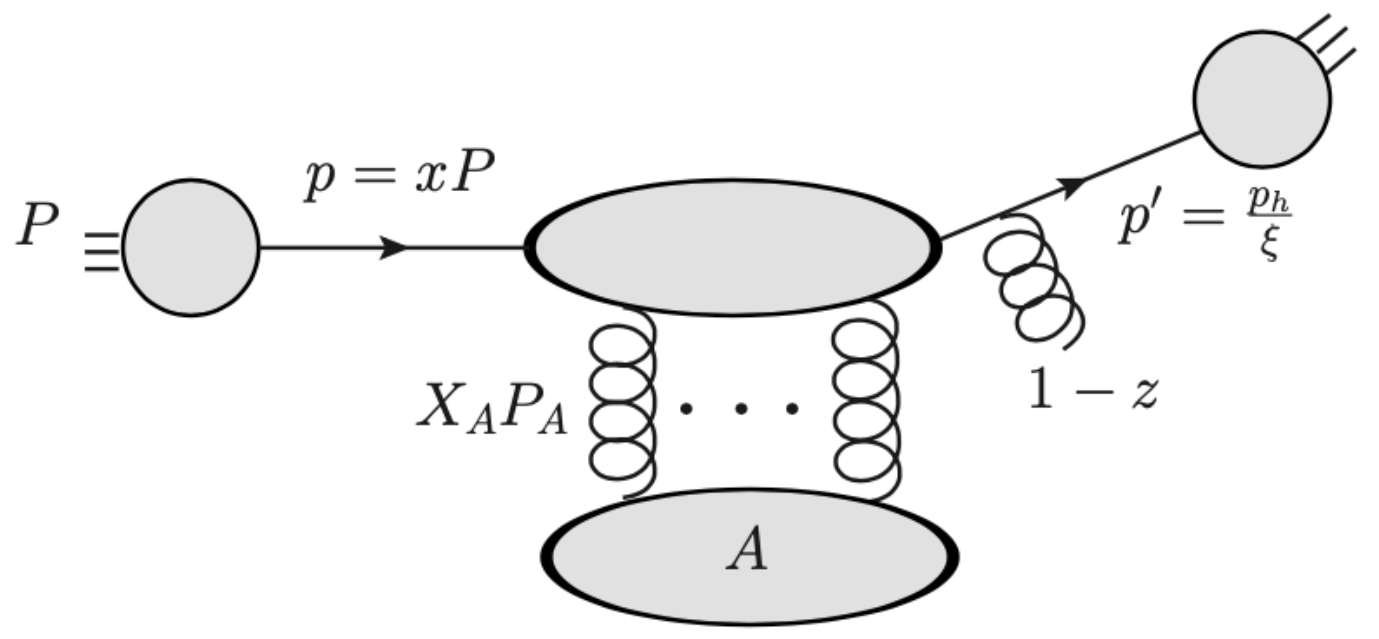} 
\caption{Illustration of $pA \to h X$.}
\label{fig:illustration}
\end{figure}

The same story happens to $pA \to hX$. The $n$-th order corrections to the partonic cross section possess the logarithmic structure in the large $N_c$ limit
\bea\label{eq:log-stru} 
\hat{\sigma}^{(n)} \supset  \sum_{k=0}^{n-1} \alpha_s^n\left( \frac{\ln^{k }(1-z)}{1-z} \right)_+
\,,
\eea 
where $1- z = 1 - \tau/x \xi$ with $x$ and $\xi$ the momentum fraction in the PDF and the fragmentation function (FF), respectively, as illustrated in fig.~\ref{fig:illustration}. Note that $1-z$ is the energy fraction carried by the bremsstrahlung radiations. We have $\tau =  p_{h,\perp} e^{y_h}/\sqrt{s}$, with $y_h$ the hadron rapidity and $p_{h,\perp}$ the transverse momentum. In the forward region, $y_h$ is very large and thus $z$ quickly approaches $1$. The system is reaching the threshold and the radiations can only be soft and the logarithms are large. 

To make it more specific, we consider the $pA \to hX$ at NLO. In the large $N_c$ limit, the partonic cross section can be written as~\cite{Chirilli:2011km,Liu:2019iml,Kang:2019ysm,futurework}
\bea\label{eq:NLO-thre}
& \frac{\mathrm{d}^2\hat{\sigma}^{(1)} }{\mathrm{d} z \mathrm{d}^2p_\perp'}
\propto - \frac{\alpha_s}{2\pi} {\bf T}^2_i   P_{i\to i}(z) \ln \frac{r_\perp^2\mu^2}{c_0^2}
  \left(1+ \frac{1}{z^2} e^{i \frac{1-z}{z}p_\perp'\cdot r_\perp} \right) \, \nn
& - \frac{\alpha_s}{\pi} {\bf T}_i^a {\bf T}_j^{a'}  
\int \frac{\mathrm{d}x_\perp}{\, \pi}
\left\{ \frac{1}{z}\tilde{P}_{i\to i}(z) 
\,  e^{i\frac{1-z}{z} p_\perp'\cdot r_\perp'} 
\frac{r_\perp' \cdot r_\perp''}{{r_\perp'}^2 {r_\perp''}^2} \right.  \nn 
& \hspace{2.ex} \left. +  \,  \delta(1-z) 
\ln \frac{X_f}{X_A}
\,
\left[ \frac{r_\perp^2}{{r_\perp'}^2 {r_\perp''}^2} \right]_+  \right\} W_{aa'}(x_\perp) 
+ \dots \,,
\eea
where we have factorized out the LO terms. At the same time, $c_0 = 2e^{-\gamma_E}$ with $\gamma_E$ the Euler constant, and $p_\perp' = p_{h,\perp}/\xi$ is the transverse momentum of the fragmenting parton. We have only written out those $(1-z)$ singular terms relevant for discussion, but suppress all the $(1-z)$ non-singular terms for simplicity. 
Here, $X_A$ is the momentum fraction carried by the gluon from the nucleus and $X_f$ is the scale due to the rapidity divergence~\cite{Fleming:2014rea,Rothstein:2016bsq,Liu:2019iml,Kang:2019ysm}. $P_{i \to i}(z)$ is the splitting function and $\tilde{P}_{i\to i}(z)$ is $P_{i \to i}(z)$ without the $\delta(1-z)$ term,  $r_\perp = b_\perp' - b_\perp$, $r_\perp' = b_\perp - x_\perp$ and $r_\perp'' = x_\perp - b_\perp'$. The $+$-prescription is defined in~\cite{Chiu:2012ir} which subtracts the singularities at $x_\perp \to b_\perp\, (b_\perp')$ and $W_{aa'}$ is the CGC Wilson line in the adjoint representation. We find it convenient to use the color operator ${\bf T}^a_i$ introduced by Catani et al.~\cite{Catani:1999ss}, acting on the $i$-th parton with color $c$($c'$) in the color space as 
\bea\label{eq:color}
\langle i_{c} \,, j_b \dots |
{\bf T}^a_{i} |i_{c'},j_{b'} \,, \dots  \rangle = 
T^a_{c,c'} \delta_{bb'}  \dots  \,, 
\eea
where $T^a_{c,b} = i f_{cab} $ if the particle $i$ is a gluon and $T^a_{c,b} = t^a_{c,b}$ for a final state quark while
$T^a_{c,b} = - t^a_{b,c}$ for a final state antiquark. 

As $z \to 1$, the splitting function $\tilde{P}_{i\to i}(z) \to \frac{2}{(1-z)_+}$ and we see explicitly in Eq.~(\ref{eq:NLO-thre}) that the NLO results reduce to the threshold structure in Eq.~(\ref{eq:log-stru}) with $n=1$ and $k=0$. After integrating over $z$, the logarithmic form will be more explicit~\cite{Sterman:1986aj,Becher:2007ty,Aicher:2010cb}. 

When $1-z \sim {\cal O}(1)$, these $(1-z)^{-1}_+$ terms are small and do no harm to the perturbative calculation. 
In this away-from-threshold case, the typical energy scales
involved are the longitudinal momentum 
 $\bar{n} \cdot p$ of the incoming parton moving along $n$ direction where 
 $n = (1,0,0,1)$ and 
 ${\bar n} = (1,0,0,-1)$, and $p_\perp'$ of the out-going parton. The heirachy $p_\perp' \ll \bar{n} \cdot p$ gives rise to large logarithms $\ln \frac{{\bar n}\cdot p}{p'_\perp}$, which we will see, can be resummed by the BK evolution, if the CGC rapidity scale choice $X_f \sim X_A$ is made.

\begin{figure}[h!]
\vspace{-6.5ex}
\hspace{-4.ex}
\includegraphics[width=\linewidth]{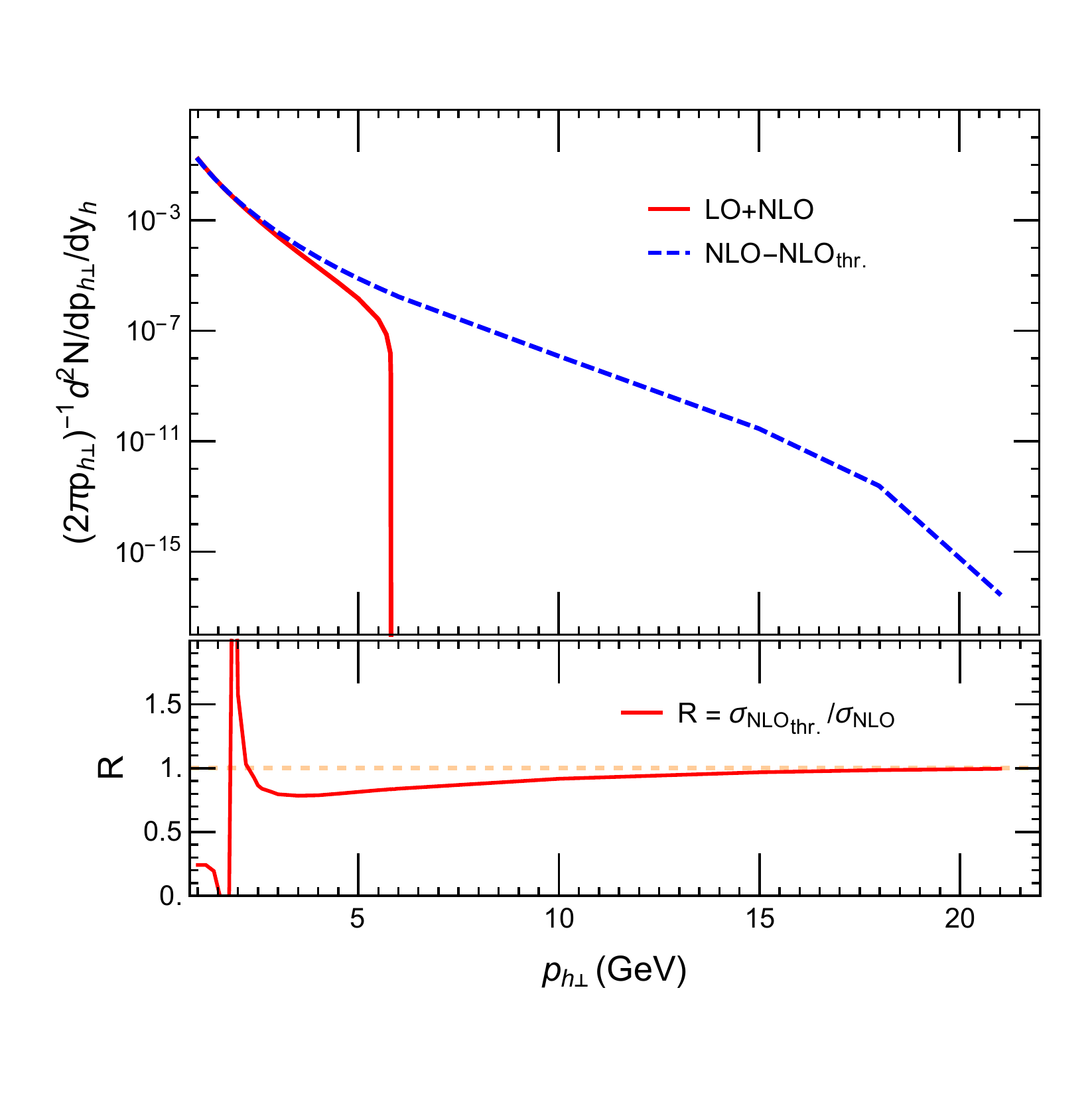} 
 \vspace{-8.5ex}
\caption{Size and the negative contribution of the threshold logarithms. %In the lower panel, common $\delta(1-z)$ contributions have been taken out in the ratio.
}
 \vspace{-4.ex}
\label{fig:threshold}
\end{figure}

However when we increase $p_{h,\perp}$, especially in the forward region where $y_h$ is large, $z$ quickly approaches its threshold and the threshold terms can become extraordinarily large. To demonstrate this point, we plot explicitly this near-threshold situation in fig.~\ref{fig:threshold},
using $dAu$ collision at RHIC with $\sqrt{s} = 200\>{\rm GeV}$ and $y_h = 2.2$ as an example. In the upper panel, the solid curve is the full NLO cross section including the kinematic constraint~\cite{Chirilli:2011km,Liu:2019iml,Kang:2019ysm}, while the dashed curve is the NLO result with the threshold $(1-z)^{-1}_+$ terms (setting $z=1$ in the numerator) in Eq.~(\ref{eq:NLO-thre}) subtracted. From this comparison, we see clearly that, when the threshold singular terms are absent, the remaining contribution stays positive for the entire $p_{h,\perp}$ spectrum, while the full NLO prediction quickly drops below zero. In the lower panel of fig.~\ref{fig:threshold}, we show the ratio $R$ between the NLO threshold contribution and the full NLO result. 
To make the plot more evident, we take out the common $\delta(1-z)$ term from
both the full NLO and the threshold contributions. 
We see that for low $p_{h,\perp}$, non-threshold terms are comparable with the threshold contributions. As $p_{h,\perp} > 5\>{\rm GeV}$, %we increase $p_{h,\perp}$, 
the threshold logarithms soon become overwhelmingly dominant %for $p_{h,\perp} > 5\>{\rm GeV}$ where the NLO cross section starts to become negative, 
and the ratio $R$ approaches one. Same behaviors are observed in all other forward kinematic settings.

This exercise clearly indicates that 1).~the threshold logarithm is the source to the negative cross sections; 2).~the threshold logarithm is enormous $\gtrsim 100\%$ of LO in magnitude and thus requires resummation. 

 {\it Away from threshold.} We start with the away-from-threshold case to introduce our formalism and notations and to highlight how large logarithms are resummed. At LO, the differential cross section within the CGC framework can be written as
\bea\label{eq:forward-LO}
&\frac{ \mathrm{d}\sigma  }{\mathrm{d}y_h \mathrm{d}^2p_{h,\perp} }
=
   \sum_{i, j = g,q}
   \frac{1}{4\pi^2} \int \frac{ \mathrm{d}  \xi }{\xi^2} \,
x_p f_{i/P}(x_p,\mu) D_{h/j}(\xi,\mu) \nn 
& \times  \int \mathrm{d}^2 b_\perp \mathrm{d}^2 b_\perp' 
 \, 
e^{ip_\perp' \cdot r_\perp } \, 
 \Big \langle 
  \langle {\bf {\cal M}}_0 (b_\perp') |{\bf {\cal M}}_0 (b_\perp) \rangle
\Big \rangle_{\nu} 
\,,
\eea
where $ \langle {\bf {\cal M}}_0 (b_\perp') |{\bf {\cal M}}_0 (b_\perp) \rangle
=\frac{1}{C} Tr[W^\dagger(b_\perp') W(b_\perp)]$, with $C = N_c$ the number of colors for quark
and $N_c^2$ for gluon initial state in large $N_c$ limit. 
We used the LO color space notation 
$|{\cal M}_0(b_\perp) \rangle$~\cite{Catani:1999ss} which includes the CGC (Glauber) Wilson line
$W_{i_cj_c}(b_\perp)$ with $i_c$ and $j_c$ the color indices for the in-coming and the out-going partons, fundamental for quark and adjoint for gluon.
$f_{i/P}$ is the PDF, $x_p = p_{h,\perp}e^{y_h}/\xi\sqrt{s}$ and $D_{h/j}$ is the FF. Here, $\nu$ is the rapidity scale~\cite{Fleming:2014rea,Rothstein:2016bsq,Kang:2019ysm} in our regularization method for the rapidity divergence in the NLO calculations, and will be related later to the gluon rapidity $Y_A\sim \ln (1/X_A)$ in the nucleus.

Beyond LO, an all-order factorization theorem can be derived using the machinery of the soft-collinear-effective theory~\cite{Rothstein:2016bsq,Bauer:2000yr,Bauer:2001yt,Bauer:2002uv,Chiu:2012ir,Becher:2015hka} with additional interactions between quarks/gluons and the Wilson line $W(x_\perp)$ adding to it~\cite{futurework}, which reads
\bea\label{eq:forward-LL}
&\frac{ \mathrm{d}\sigma  }{\mathrm{d}y_h \mathrm{d}^2p_{h\perp} }
=
   \sum_{i, j = g,q}
   \frac{1}{4\pi^2} \int \frac{ \mathrm{d}  \xi }{\xi^2} \,
\frac{\mathrm{d} x}{x}\,
z x f_{i/P}(x,\mu) D_{h/j}(\xi,\mu) \nn 
& \times  \int \mathrm{d}^2 b_\perp \mathrm{d}^2 b_\perp' 
 \, 
e^{ip_\perp' \cdot r_\perp } \, 
 \\ 
&  \hspace{-2.ex}  \times  \Big \langle 
  \langle {\bf {\cal M}}_0 (b_\perp') | \, 
\bm{{\cal J}}(z,\mu,\nu,b_\perp,b_\perp')  
\bm{{\cal S}}(\mu,\nu,b_\perp,b_\perp')|{\bf {\cal M}}_0 (b_\perp) \rangle
\Big \rangle_\nu 
\,. \nonumber
\eea
Here the collinear function $\bm{{\cal J}}$ involves the leading power SCET collinear fields~\cite{Rothstein:2016bsq} and encodes the corrections from radiations with the momentum scaling as $({\bar n}\cdot p, n \cdot p, p_\perp) \sim
\sqrt{s} (1,\lambda^2, \lambda)$, while the soft function $\bm{{\cal S}}$ is made up of the soft Wilson lines of the soft gluons with the momentum scaling $k \sim
\sqrt{s} (\lambda,\lambda, \lambda)$. To reach the factorization, the standard field redefinition following~\cite{Bauer:2001yt} is performed to factorize the soft and collinear contributions.
The derivation is a bit similar to~\cite{Becher:2015hka} which deals with the non-global logarithms and will be presented in~\cite{futurework}. 
The collinear and soft sectors are classified using the observable power counting in~\cite{Kang:2019ysm} and can be calculated perturbatively. At the LO, $\bm{{\cal J}}(z) = {\bm 1} \delta(1-z)$ and $\bm{{\cal S}}={\bm 1}$ and we reproduce Eq.~(\ref{eq:forward-LO}).
Beyond LO, dimensional regularization and additional rapidity regularization are required
to regulate the divergences in the collinear and the soft function, which generates the $\epsilon$ and $\eta$ poles and
the collinear scale $\mu$ and the rapidity scale $\nu$ dependence~\cite{Liu:2019iml,Kang:2019ysm}.

With the scale choice $\mu \sim p_{h,\perp}$, all logarithms involving the scale $\mu$ are minimized and absorbed into the evolutions of PDFs/FFs. Hence we only focus on the logarithms associated with the scale $\nu$. To all orders, $\bm{{\cal J}}$ and $\bm{{\cal S}}$  satisfy the rapidity renormalization group equations
\bea\label{eq:rge}
\nu \frac{\mathrm{d}  }{\mathrm{d} \nu} \, \bm{{\cal F}}(\nu)
= \kappa\,  \bm{\gamma}_\nu \, \bm{{\cal F}}(\nu)  \,,
\eea 
where $\bm{{\cal F}}=\bm{{\cal J}}$ or $\bm{{\cal S}}$. The rapidity anomalous dimension $\kappa \, \bm{\gamma}_\nu$ can be read off from the $\eta$-poles in the soft and the collinear functions, %in perturbative calculations. 
which is calculated at NLO in~\cite{Kang:2019ysm, futurework} to find
\bea
\bm{\gamma}_{\nu} =  -  \, \frac{\alpha_s}{\pi}  \, 
 \int  \frac{\mathrm{d}x_\perp }{\pi}\,
\left[ \frac{{r_\perp}^2}{{r_\perp'}^2 {r_\perp''}^2} \right]_+ 
 {\bf T}^a_i  {\bf T}^{a'}_j  \, W_{aa'}(x_\perp )  \,, 
\eea
 with $\kappa = -1 (2) $ for $\bm{{\cal J}}$($\bm{{\cal S}}$). %the collinear (soft) function. 
Here $[\dots]_+$ is the %well-known 
BK evolution kernel, denoted as $I_{BK}$ below. We can solve Eq.~(\ref{eq:rge}) %the renormalization group equation 
to find $\bm{{\cal F}}(\nu) =  {\bm U}_{\cal F}(\nu,\nu_{\cal F}) \, \bm{{\cal F}} (\nu_{{\cal F}})$, and the evolution kernel ${\bm U}_{\cal F}$ 
 evolves both functions from their natural scale $\nu_{\cal F}$ to a common scale $\nu$ to evaluate the cross section meanwhile resums large logarithms $\ln \frac{\nu}{\nu_{\cal F}}$. The $\nu_{\cal F}$ is determined by minimizing the logarithms in $\bm{{\cal F}}$  
 and leads to
$ \nu_J = {\bar n}\cdot p \,,
\nu_S = p_\perp'$  
for the collinear and the soft sectors~\cite{Kang:2019ysm}. 
At LL, we find  
 \bea\label{eq:evolution-non-thr}
 {\bm U}_{J} {\bm U}_{S}
 = \exp\left[  \,  \bm{\gamma}_{\nu} \ln \frac{\nu \, \nu_J}{\nu_S^2}
 \right] 
 =   \exp\left[  \,  \bm{\gamma}_{\nu} \ln \frac{X_f}{X_A}
 \right] 
 \,,
 \eea 
  which resums large logarithms of the form $\ln \frac{\nu}{{\bar n}\cdot p}$ and $\ln \frac{\nu}{p_\perp'}$ in $\bm{{\cal J}}$ and $\bm{{\cal S}}$, respectively. 
%the collinear and the soft functions, respectively. 
Here we have used $\nu/( \nu_S^2/\nu_J) = \nu/({p_\perp'}^2/{\bar n}\cdot p) = X_f/X_A$, where $X_f = \nu/n\cdot P_A$ and $X_A =
\frac{{p_\perp'}^2}{{\bar n}\cdot p n\cdot P_A}
$ with $P_A$ the momentum of the nucleus, to get the second equation.

The $\nu$-independence of the cross section implies the evolution 
for the dipole $W^\dagger(b_\perp') \otimes W(b_\perp)$
 \bea
&   \nu\frac{ \mathrm{d} }{\mathrm{d} \nu} W^\dagger_{j'_c, i'_c}(b_\perp')
W_{i_c\, j_c}(b_\perp)  
\,
 =   \, \frac{\alpha_s}{\pi}  \, 
 \int  \frac{\mathrm{d}x_\perp }{\pi}\,
\left[ \frac{{r_\perp}^2}{{r_\perp'}^2 {r_\perp''}^2} \right]_+  \nn 
&\hspace{1.5ex} \times 
\left[ T^a 
 W^\dagger(b_\perp')\right]_{j'_c, i'_c}
\, \left[
 T^{a'}
 \, W(b_\perp) \right]_{i_c\, j_c}\,   W_{aa'}(x_\perp ) \,,
 \eea
 which when traced over, is nothing but the BK equation. %Here the color notation $T^a_{b,c}$ is defined previously in Eq.~(\ref{eq:color}) and not to be confused with the fundamental representation $t^a_{b,c}$.
 
With the evolution in Eq.~(\ref{eq:evolution-non-thr}), the choice of the rapidity scale $X_f$ (or equivalently $\nu$) could in principle be arbitrary, since all large logarithms are resummed. One natural choice is to set $X_f = X_A$ which is nothing but the conventional CGC scale choice. In such a way, one only needs to evolve the CGC dipoles $W^\dagger \otimes W$ since the evolution ${\bm U}_J{\bm U}_S = 1$. %gets eliminated. 
In other words, all large logarithms $\ln \frac{p_\perp'}{{\bar n}\cdot p}$ are effectively absorbed into the dipole evolution, if $X_f \sim X_A$, when away from threshold.

 {\it Near threshold}.  
 When near the threshold, real energetic collinear radiations are forbidden, since the longitudinal momentum of the radiation ${\bar n} \cdot p(1-z) $ is restricted to be soft as $z \to 1$, while virtual collinear corrections are still allowed~\cite{Kang:2019ysm}. Therefore, in the threshold limit, the collinear function $\bm{{\cal J}}_{thr.}$ only contains the collinear virtual corrections. All real radiations are now soft and encoded in $\bm{{\cal S}}_{thr.}$. In the large $N_c$ limit, it is found that still only the soft and collinear modes contribute at the leading power~\cite{Kang:2019ysm} and the form of the factorization theorem remains the same as Eq.~(\ref{eq:forward-LL}) but with the replacement $\bm{{\cal J}}(z) \to \bm{{\cal J}}_{thr.}$ and $\bm{{\cal S}} \to \bm{{\cal S}}_{thr.}(z)$. 
 
 The NLO $\bm{{\cal J}}_{thr.}$ is exactly the NLO virtual corrections of $\bm{{\cal J}}$, which gives the evolution
  \bea
 {\bm U}_{{J}_{thr.}} 
 = \exp\left[  \,   
 \frac{\alpha_s }{\pi } \ln \frac{\nu}{\nu_J}
 \int  \frac{\mathrm{d}x_\perp }{\pi}
I_{BK,v}
 {\bf T}^a_i  {\bf T}^{a'}_j  W_{aa'}(x_\perp ) 
 \right],\nonumber
 \eea
 where $I_{BK,v}(r_\perp') = \left[ \frac{e^{ip_\perp'\cdot r_\perp'}}{{r_\perp'}^2} 
 \right]_+ e^{-ip_\perp' \cdot r_\perp' }$, 
 is the NLO virtual correction to the BK kernel and $\nu_J \sim {\bar n}\cdot p$ to avoid the occurrence of the large logarithms within $\bm{{\cal J}}_{thr.}$. 
 
 %On the other hand
 The calculation of the NLO threshold soft function is depicted in~\cite{Kang:2019ysm}, which gives
 \bea
 \bm{{\cal S}}_{thr.}(\nu_S) =&  \delta(1-z) 
 \left[{\bm 1} 
   + 
   \bm{{\cal S}}^{(1)}
 \right]  - \frac{\alpha_s\, {\bf T}_i^2 }{\pi} 
 \frac{2}{(1-z)_+} 
% \ln \frac{r_\perp^2\mu^2}{c_0^2}
 \ln \frac{\mu^2}{{p_\perp'}^2}
   \nn
&\hspace{-23mm}+  \frac{\alpha_s}{\pi} \,{\bf T}_i^a {\bf T}_j^{a'}
\left[ \frac{\, \nu_S}{{\bar n} \cdot p \,  (1-z) } \right]_+
\int \frac{\mathrm{d}x_\perp}{\pi}
 I_{BK,r} \,  W_{aa'}(x_\perp), 
%  \nn
%&
%- \frac{\alpha_s\, {\bf T}_i^2 }{\pi}
%   \frac{2}{(1-z)_+}  
%  \ln \frac{r_\perp^2{p_\perp'}^2}{c_0^2}
\label{eq:threshold-soft}
 \eea
 where $I_{BK,r} = I_{BK} - I_{BK,v}$ is the real contribution to the BK evolution kernel. 
Here $ \bm{{\cal S}}^{(1)}$ is the NLO soft function for the away-from-threshold case~\cite{Kang:2019ysm, futurework}, which contains the kinematic constraints. The second term got its contribution from the initial and final parton splitting, which will be absorbed into the threshold evolution of the PDF/FFs and this contribution has been considered in~\cite{Xiao:2018zxf}. However, we note that this term alone is not responsible for the negative contribution and therefore its resummation can not resolve the negative cross section problem.
We can perform the Mellin transformation $\int \mathrm{d}z z^{N-1}  \bm{{\cal S}}_{thr.}(z)$ to the soft function to find $\nu_S \sim p_\perp' \sim  \frac{ {\bar n}\cdot p }{N e^{-\gamma_E}}$ which minimizes the logarithms in $ \bm{{\cal S}}_{thr.}$. 
We find the associated evolution gives
 \bea
 {\bm U}_{{S}_{thr.}}
 = & \exp\left[  \,   
 \frac{\alpha_s }{\pi } \ln \frac{\nu}{\nu_S}
 \int  \frac{\mathrm{d}x_\perp }{\pi} 
 \left( I_{BK,r } -2 
\left[ \frac{{r_\perp}^2}{{r_\perp'}^2 {r_\perp''}^2} \right]_+ 
\right) \right. \nn
& \left. \times
 {\bf T}^a_i  {\bf T}^{a'}_j  \, W_{aa'}(x_\perp ) 
 \right] \,.
 \eea
 We merge both the evolutions to find
 \bea\label{eq:threshevo}
 {\bm U}_{{J}_{thr.}} & {\bm U}_{{S}_{thr.}}
=   \exp\left[  \, - 
 \frac{\alpha_s }{ \pi }  \int  \frac{\mathrm{d}x_\perp }{\pi}
 \left(  \ln \frac{\nu_S}{\nu_J} I_{BK,r}  \right. \right. \nn
& \left. \left. + \ln \frac{X_f}{X_A}   
I_{BK}
\right) 
   {\bf T}^a_i  {\bf T}^{a'}_j  \, W_{aa'}(x_\perp ) 
 \right]   \,,
 \eea 
 where we notice that the second term is identical to the away-from-threshold evolution while the additional first term arises to resum the threshold logarithms. The probability for emitting a soft parton (real correction) is suppressed after resummation. %The results could also be realized by considering strongly rapidity ordered soft emissions~\cite{futurework} and are extensible to other small-$x$ processes. 

 From the result, we see that, when near threshold, suppose we still stick to the %conventional CGC 
scale choice $X_f = X_A$, then there requires an additional evolution factor to account for the threshold impacts not covered by simply evolving the CGC dipole. 
 
Given that a different rapidity scale choice
$X_f$ in the nuclear target will be compensated by the corresponding evolution factor in
Eq.~(\ref{eq:threshevo}), %In principle, once resummed, 
the result is ignorant of the $X_f$ choice. Therefore, instead, we can dynamically determine $X_f$ by demanding it minimizing the exponent in Eq.~(\ref{eq:threshevo}) following the similar procedure in~\cite{Becher:2009th,Becher:2011fc}, and hence eliminate the complicated evolution but still maintain the threshold resummation to all orders. The idea is similar to set $X_f \sim X_A$ in the away-from-threshold case. We will use this approach for phenomenology studies. 
 
 {\it Phenomenology.} 
Now we illustrate the numerical NLO$+$LL$_{\rm thr.}$ predictions for the kinematics relevant to both the RHIC and LHC experiments. We include all partonic channels. We used MSTW2008 PDF sets~\cite{Martin:2009iq} and DSS parametrizations~\cite{deFlorian:2014xna,deFlorian:2017lwf} for the FFs. The CGC dipoles are obtained by solving the LL BK equation with the running coupling correction~\cite{Kovchegov:2006wf,Kovchegov:2006vj,Balitsky:2006wa}, with the parameters used in~\cite{Fujii:2013gxa}. We set the collinear factorization scale $\mu = p_{h,\perp}$. For fixed kinematics, we determine the central rapidity scale by scanning through $X_f$ (or equivalently $\nu$) numerically to find the value that minimizes the exponent in Eq.~(\ref{eq:threshevo}).%, which accounts for the threshold LL resummation.
 \begin{figure}[h!]
\vspace{-1.5ex}
\hspace{-5.ex}
\includegraphics[width=\linewidth]{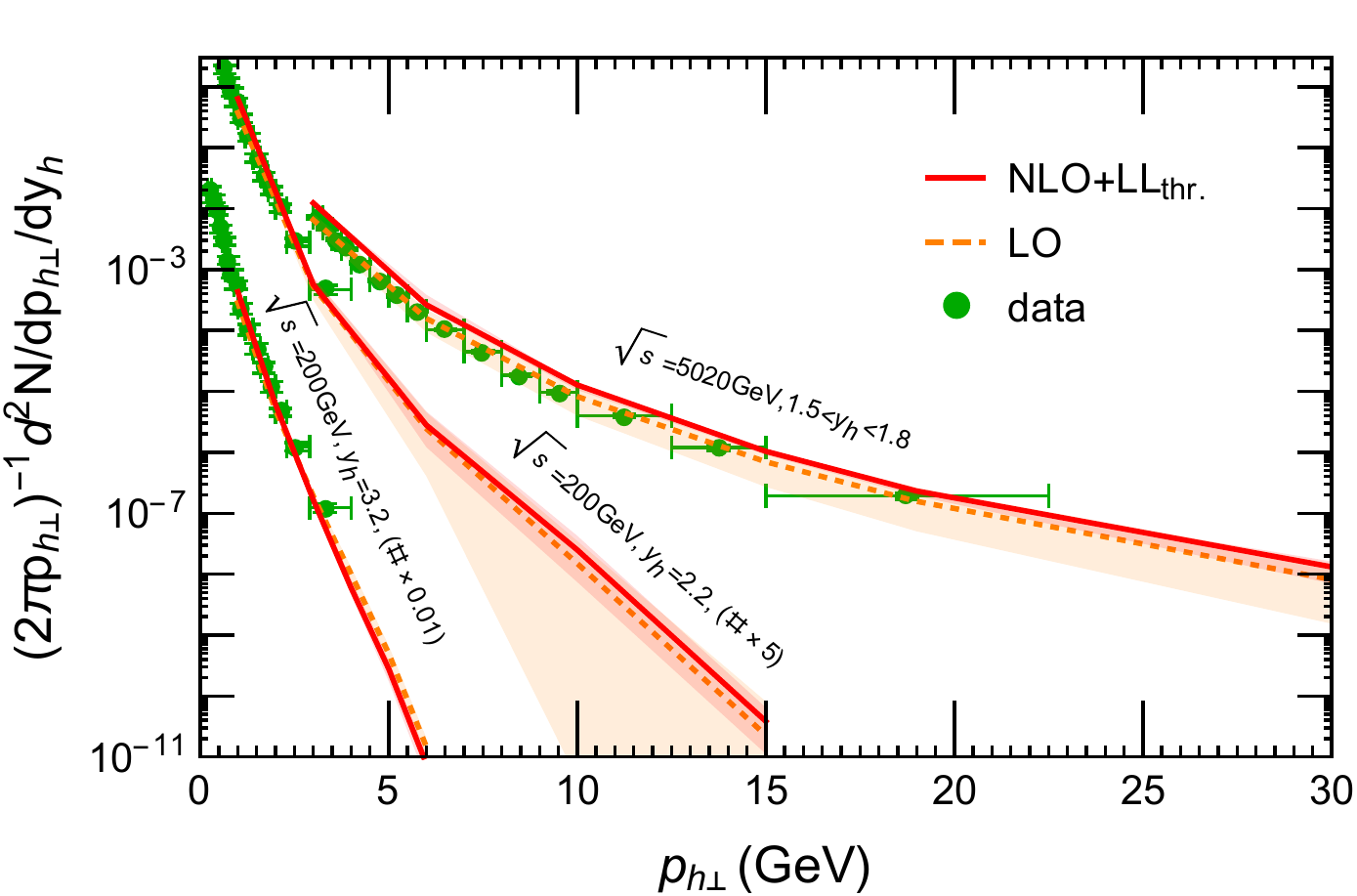} 
 \vspace{-1.6ex}
\caption{Data versus theory predictions.}
 \vspace{-5.ex}
\label{fig:threshold-data}
\end{figure}

We present the predictions in fig.~\ref{fig:threshold-data}, where we compare the theoretical results with the data in the forward rapidity region from the charged hadron production in p+Pb collisions at LHC %with $\sqrt{s} = 5.02 \>{\rm TeV}$, $1.5<y_h<1.8$~\cite{Aad:2016zif} 
and the %negatively charged 
hadron productions in d+Au collisions at 
RHIC%with $\sqrt{s} = 200\>{\rm GeV}$, $y_h = 2.2$ and $y_h = 3.2$
~\cite{Arsene:2004ux}. From fig.~\ref{fig:threshold-data}, we see that the NLO$+$LL$_{\rm thr.}$ results stay positive %all the way to large $p_{h,\perp}$ 
and show no signs of turning negative. The uncertainty bands %represent the rapidity scale $X_f$ variation, which 
are obtained by varying $X_f$ around its central value up and down by a factor of~$2$ and taking the maximum deviations. We see that the uncertainties are substantially reduced when we go from LO (orange bands) to NLO$+$LL$_{\rm thr.}$ (red bands). 
The NLO$+$LL$_{\rm thr.}$ calculation %does an impressive job in 
impressively describes all the experimental data. %(green dots in fig.~\ref{fig:threshold-data}). 
The central values of the predictions slightly overshoot the LHC data for small $p_{h,\perp}$ but still within errors. The situation is expected to be further improved if a global
fit beyond LO is performed to determine the CGC dipole initial condition.

{\it Conclusions.} In this paper, through thorough studies, we identify the threshold logarithms responsible for the negative cross section problem that are missing in previous discussions~\cite{Xiao:2018zxf} in the forward %hadron production in %proton-nucleus collisions, 
$pA\to hX$, within the small-$x$ formalism. We develop an all-order factorization theorem with systematically improvable accuracy. We present detailed derivation and numerical study for the first complete threshold resummation at LL in the CGC formalism. We find that the LL$_{\rm thr.}$ resummation can be realized simply by a suitable rapidity scale choice in the NLO calculation. After resummation, all predicted $p_{h,\perp}$ spectrums are found to be positive all the way to the kinematic boundaries. We compared our predictions with the available data and observed excellent agreements with greatly reduced scale uncertainties, in comparison with the LO results. Our results are ready for more phenomenological applications at the LHC and RHIC, such as global fitting studies of the CGC models beyond LO. Given the universality of the LL$_{\rm thr.}$ structure in hadronic processes, we expect our approach is applicable to many other practical applications of high order CGC predictions for the small-$x$ collider phenomenology.
 
{\it Acknowledgements.} We thank Yan-Qing Ma, Felix Ringer, Feng Yuan and Mao Zeng for valuable discussions and comments on the draft. This work is supported by the National Science Foundation under Contract No.~PHY-1720486 (Z.K.) and by the National Natural Science Foundation of China under Grant No.~11775023 (X.L.). 

%%%%%%%%%%

\end{document}